# Mechanics and Tunable Bandgap by Straining in Single-Layer Hexagonal Boron-Nitride

Jiangtao Wu[1], Baolin Wang[1], Yujie Wei[1*], Ronggui Yang[2], Mildred Dresselhaus[3]

[1]LNM, Institute of Mechanics, Chinese Academy of Sciences, Beijing, 100190, People's Republic of China

[2]Department of Mechanical Engineering, University of Colorado, Boulder, CO 80309, USA

[3]Departments of Physics and Department of Electrical Engineering and Computer Sciences, Massachusetts Institute of Technology, Cambridge, MA 02139, USA

*Correspondence: Y.Wei, Phone: (86-10) 8254-4169; email: yujie_wei@lnm.imech.ac.cn,

# ABSTRACT

Current interest in two-dimensional materials extends from graphene to others systems like single-layer hexagonal boron-nitride (h-BN), for the possibility of making heterogeneous structures to achieve exceptional properties that cannot be realized in graphene. The electrically insulating h-BN and semi-metal graphene may open good opportunities to realize a semiconductor by manipulating the morphology and composition of such heterogeneous structures. Here we report the mechanical properties of h-BN and its band structures tuned by mechanical straining by using the density functional theory calculations. The elastic properties, both the Young's modulus and bending rigidity for h-BN, are isotropic. However, its failure strength and failure strain show strong anisotropy. We reveal that there is a bi-linear dependence of band gap on the applied tensile strains in h-BN. Mechanical strain can tune single-layer h-BN from an insulator to a semiconductor, with a band gap in the 4.7eV to 1.5eV range.

As another member of the two-dimensional (2D) material family, single layer hexagonal boron-nitride (h-BN) [1-8] has received increasing attention recently, due to the great similarity between h-BN and graphene and the potential to be heterogeneously integrated with graphene electronics. Graphene and h-BN resemble one another in many aspects, e.g., the capability to form tubes [2,9], a honeycomb arrangement in *sp²*-bonded 2D layers, high thermal conductivity and excellent lubrication properties. Furthermore, they also exhibit significant differences: h-BN is electrically insulating with a large band gap both within and across the layers, in contrast to its rival graphene with no band gap and being a semi-metal [10-12]. Such striking differences in physical properties stimulate great interest in the heterogeneous integration of both materials for multifunctional applications [6,10,13]. For example, there is growing interest to tune the band gaps of two-dimensional materials by adopting hybrid structures to utilize the large band gap of h-BN and the zero band gap of graphene to potentially realize a target band gap. This justifies the course to re-examine h-BN for all the extraordinary properties observed in graphene. For example, the mechanical properties of h-BN tubes and the defects in h-BN sheets have been extensively investigated [7,14-17]. However, a thorough investigation of the mechanical properties of single layer h-BN and how mechanics may influence the band structures of h-BN are still missing. In this work, we perform a systematic investigation on several critical mechanical properties of single-layer h-BN, including the Young's modulus, the Poisson ratio, bending rigidity and the band structure tunability by mechanical strain.

First principles density functional theory (DFT) calculations on single layer h-BN were performed (see Appendix for details) and are reported here. Following the rules to define the chirality of carbon nanotubes and graphene, the orientation of a single layer crystalline h-BN is described by two vectors $\boldsymbol{a}_1$ and $\boldsymbol{a}_2$, and h-BN has a chiral vector $\boldsymbol{C}_h = n\boldsymbol{a}_1 + m\boldsymbol{a}_2$, as illustrated in Fig. 1a. Mechanical loading is applied to four different orientations in h-BN, as defined by chiralities: (I) zigzag $C_h$=(1,0); (II) armchair $C_h$=(1,1); (III) $C_h$=(4,1); and (IV) $C_h$=(2,1), such that $\theta$, which is the angle between the loading direction (vector $C_h$) and the zigzag direction ($a_1$), changes from 0° to 30° with an incremental of about 10°. In the simulations, we first fully relax the unit cell to determine the equilibrium lattice constant and a lattice constant of $a$=2.504Å is found, which agrees well with the experimental measurement of 2.506Å [18]. Except for the bi-axial loading case, we relax the box in the direction perpendicular to the loading axis in the sheet plane in all other calculations. To obtain the equivalent stress of the single layer h-BN, an interlayer distance $h$=3.3Å [18-23] is used, following reference [24].

Fig. 1b shows the engineering stress versus engineering strain curves for the samples subjected to loading in the four directions defined by chiral vectors $C_h$, as mentioned above. In all cases, h-BN exhibits linear elastic response until a strain of about 8%, followed by a nonlinear elastic behavior until its strength reaches a plateau. Table 1 summarizes the Young's modulus, the strength, the failure strain corresponding to the point at the strength, as well as the critical bond length at failure.

Similar to the mechanical properties of graphene [24, 25], we find that the strength and the failure strain in h-BN are strongly anisotropic. The strength and failure strain when pulling along the zigzag direction is much higher than that along the armchair direction. This can be well explained by the honeycomb $sp^2$-bonded atomic arrangement. The maximum distance of two neighboring planes parallel to the zigzag direction is $d_{ZZ}=a/\sqrt{3}$=1.446Å, while the distance between atomic planes parallel to the armchair direction is $d_{AC}=a/2$=1.252Å, where $a$=2.504Å is the lattice constant of h-BN. Fig. 1c shows the Poisson's ratio $\nu$ as a function of the tensile strain applied. Compared to the isotropic Young's modulus, Poisson's ratio is anisotropic. The Poisson's ratio is defined as $v = -\frac{L_T - L0_T}{L_{Ch} - L0_{Ch}} \frac{L0_{Ch}}{L0_T}$, where $L0_T$, and $L0_{Ch}$ are the initial lengths of the unit cell in the directions perpendicular and parallel to the loading axis, respectively, while $L_T$, and $L_{Ch}$ are the corresponding ones after an applied strain. As seen in Fig. 1c, the Poisson's ratio under small strain increases with increasing strain. However, when a larger strain of more than 3% is applied, the Poisson's ratio decreases almost linearly with the tensile strain. We further notice that the change in the Poisson's ratio by straining in the zigzag direction is more significant than that by straining in the armchair direction. With $\theta$ increasing from 0° to 30° (the chiral vector changes from zigzag to armchair), the Poisson's ratio decreases from a value of 0.23 in the zigzag direction to 0.17 at 20% strain while that in the armchair direction changes from 0.20 to about 0.09.

The correlations among the bond angle, bond stretch and the loading conditions at the failure point where the stress is maximal are also closely investigated in the present work. Fig. 2 shows the bond lengths and bond angles of the h-BN structure at the maximum stress. For straining along the armchair direction where the critical strain is the smallest among all the cases studied in this work, the bond opening angle of 134.59° is the maximum (Fig. 2a). While the largest critical strain can be achieved when pulling along the zigzag direction (Fig. 2b), the maximum bond stretch of 1.821Å is observed for the biaxial strain case (Fig. 2c). The bond lengths and bond angles at the failure point are shown in Fig. 2d and 2e when single-layer h-BN is stretched along the orientations of $\boldsymbol{C}_h = (4,1)$ and $\boldsymbol{C}_h = (2,1)$, respectively.

Two-dimensional h-BN and graphene not only have a high modulus and tensile strength but also have extremely small out-of-plane stiffness given the material is the ultimate thin membrane. Such a combination of mechanical properties makes them ideal candidates for biological membranes and stretchable electronics applications. Here we also studied the bending rigidity $B_M$ of single layer h-BN, which governs the morphology of two-dimensional materials under external field stimuli. To obtain $B_M$ of single-layered h-BN, we perform DFT calculations to get the energies of single wall h-BN nanotubes (SWBNT) of different radii. By using the Helfrich Hamiltonian [26], the bending rigidity of a free-standing single-layer h-BN is then connected with the energy of a B-N pair $E_{BN}$ in single wall h-BN nanotubes with the radii $r$ [26]:

$$E_{BN} = E_0 + S_0 B_M r^{-2}/2 \qquad (1),$$

where $E_0$ is the energy of one B-N pair in a flat h-BN sheet, and $S_0 = 3\sqrt{3}d^2/2 = 5.432\text{Å}^2$ is the planar footprint of a B-N pair where the B-N bond length is $d = 1.446$Å. In the absence of deformation, $E_0$ for single-layer h-BN is calculated to be -17.568eV using DFT. The dots in Fig. 3 are the energies per B-N pair for different radii in single wall BN nanotubes based on DFT calculations, with 3a to 3c corresponding to zigzag tubes, armchair tubes, and tubes rolled up along the $C_h$=(5,2) direction. The solid line in Fig. 3 shows the prediction of Eq. (1), assuming a bending rigidity $B_M$ of 0.95eV. Surprisingly, a single value of bending rigidity $B_M = 0.95$eV in eq. (1) can describe very well the dependence of $E_{BN}$ on tube radii $r$, regardless of the roll-up direction of the h-BN tubes, indicating the isotropic nature of the bending rigidity for single layer h-BN, which is similar to the bending rigidity for single layer graphene.

The zero band gap of graphene and the insulating behavior of h-BN have rendered great interest in finding effective ways to tune the band gap of 2-D materials [27]. For example, it is known that mechanical straining can alter the band gaps of graphene nanoribbons significantly [28-34]. Similarly, it has also been shown that straining can change the band gaps for h-BN nanoribbons [35] and large area h-BN [36]. We pointed out that in [36], the authors used a very different loading condition and reported that the bandgap can eventually approach zero. Those results contradict with calculations for h-BN nanoribbons [35]. Here we explore the dependence of the band gap on mechanical straining in large h-BN samples. Fig. 4a shows the two-atom unit cell used for band structure calculations with Fig. 4b showing the corresponding reciprocal lattice and the first Brillouin zone. Figure 4c shows the band structure of single layer h-BN in the absence of strain. An indirect band gap of 4.7eV is obtained, which is in good agreement with the former DFT calculation result obtained using the local density approximation (LDA) [37,38], but lower than that from calculations obtained by using the quasiparticle GW approximation (GWA) (5.95eV) [39]. Fig. 4d–4f, in turn, show the evolution of band structures of single layer h-BN when uni-axial strains are exerted along the zigzag direction, along the armchair direction, and bi-axial strains are exerted. Very different from graphene, which has been shown that mechanical strain has very litter effect on the band structure of large-area graphene [40,41], the band gap of h-BN decreases dramatically with increasing tensile strain. Fig. 5 shows the band gap as a function of different values of the strain. There is a bi-linear dependence of band gap on strains: the band gap decreases slowly with the increasing strain when the applied strain is smaller than 18%; but showing a quicker decrease when a strain larger than 18% is applied. Such a dependence of band gap on straining is also seen to be anisotropic: the band gap is more sensitive to applied bi-axial strains than to the two applied uni-axial strains. Mechanical strain can tune single-layer h-BN from an insulator to a semiconductor, with a band gap in the 4.7eV to 1.5eV range.

In summary, we investigated several critical mechanical properties of single-layered h-BN by using DFT calculations. We find that the Young's modulus of single-layered h-BN is about 780GPa and is nearly independent of crystalline orientation. The bending rigidity is also isotropic and is about 0.95eV. Nevertheless, the failure behavior and the Poisson's ratio of h-BN are highly anisotropic. In addition, we investigated the band structures of single-layered h-BN as a function of different mechanical straining, and we found that there is a bi-linear dependence of band gap on the applied tensile strains. Mechanical strain can tune single-layer h-BN from an insulator to a semiconductor, with a band gap in the range from 4.7eV to 1.5eV.

**Appendix**

The DFT calculations are performed with the Vienna Ab initio Simulation Package (VASP) code [42,43]. The projector augmented wave (PAW) pseudopotentials [44,45] and the generalized gradient approximation (GGA) of the Perdew-Burke-Ernzerhof (PBE) functional [46,47] are used. A plane-wave basis set with a kinetic-energy cut-off of 400 eV and a Monkhorst-Pack [48] k-point mesh of 31×31×1 ($\Gamma$ included) are used for static electronic structure calculations. 101 uniformly spaced k-points between two high symmetrical points are used in the non-self-consistent calculation to obtain the band structure. An M-P k-point mesh of 1×1×15 was used for the structure of single-wall h-BN tubes with chiral index (*n*=5, *m*=0) in the mechanical property calculations. To eliminate the interactions between periodic images of h-BN sheets a vacuum space of 20Å was used. Periodic boundary conditions are applied to the two in-plane directions in all the calculations conducted here. All structures are relaxed using a conjugate gradient algorithm until the atomic forces are converged to 0.01eV/Å.

**ACKNOWLEDGMENT**

Y.W. acknowledges the support from Chinese Academy of Sciences (CAS), MOST 973 of China (Nr. 2012CB937500), and National Natural Science Foundation of China (NSFC) (11021262, 11272327). R.Y. acknowledges the support from AFOSR (Grant No. FA9550-11-1-0109). M.S.D. acknowledges the upport of the Solid State Solar-Thermal Energy Conversion Center (S3TEC), an Energy Frontier Research Center funded by the U.S. Department of Energy, Office of Science, Office of Basic Energy Sciences under Award Number: DE-SC0001299/DE-FG02-09ER46577. All calculations are performed at the Supercomputing Center of CAS.

**REFERENCES**

[1] K. S. Novoselov, D. Jiang, F. Schedin, T. J. Booth, V. V. Khotkevich, S. V. Morozov, and A. K. Geim, Proc. Natl. Acad. Sci. USA **102**, 10451 (2005).
[2] N. G. Chopra, R. J. Luyken, K. Cherrey, V. H. Crespi, M. L. Cohen, S. G. Louie, and A. Zettl, Science **269**, 966 (1995).
[3] D. Pacilé, J. C. Meyer, C. O. Girit, and A. Zettl, Appl. Phys. Lett. **92**, 133107 (2008).
[4] J. C. Meyer, A. Chuvilin, G. Algara-Siller, J. Biskupek, and U. Kaiser, Nano Lett. **9**, 2683 (2009).
[5] C. Jin, F. Lin, K. Suenaga, and S. Iijima, Phys. Rev. Lett. **102** 195505 (2009).
[6] N. Alem, R. Erni, C. Kisielowski, M. Rossell, W. Gannett, and A. Zettl, Phys. Rev. B **80,** 155425 (2009).
[7] L. Song *et al.*, Nano Lett. **10**, 3209 (2010).
[8] P. Sutter, R. Cortes, J. Lahiri, and E. Sutter, Nano Lett. **12**, 4869 (2012).
[9] S. Iijima, Nature **354**, 56 (1991).
[10] K. Watanabe, T. Taniguchi, and H. Kanda, Nat. Mater. **3**, 404 (2004).
[11] K. S. Novoselov, A. K. Geim, S. V. Morozov, D. Jiang, Y. Zhang, S. V. Dubonos, I. V. Grigorieva, and A. A. Firsov, Science **306**, 666 (2004).
[12] K. S. Novoselov, A. K. Geim, S. V. Morozov, D. Jiang, M. I. Katsnelson, I. V. Grigorieva, S. V. Dubonos, and A. A. Firsov, Nature **438**, 197 (2005).
[13] Y. Kubota, K. Watanabe, O. Tsuda, and T. Taniguchi, Science **317**, 932 (2007).
[14] D. Golberg, Y. Bando, Y. Huang, T. Terao, M. Mitome, C. Tang, and C. Zhi, ACS Nano **4**, 2979 (2010).
[15] N. G. Chopra, and A. Zettl, Solid State Commun **105**, 297 (1998).
[16] H. Bettinger, T. Dumitrică, G. Scuseria, and B. Yakobson, Phys. Rev. B **65,** 041406 (2002).
[17] Y. Liu, X. Zou, and B. I. Yakobson, ACS Nano **6**, 7053 (2012).

[18] W. Paszkowicz, J. B. Pelka, M. Knapp, T. Szyszko, and S. Podsiadlo, Appl. Phys. A **75**, 431 (2002).
[19] Y. Shi *et al.*, Nano Lett. **10**, 4134 (2010).
[20] Y. Lin, T. V. Williams, and J. W. Connell, J. Phys. Chem. Lett. **1**, 277 (2009).
[21] V. L. Solozhenko, G. Will, and F. Elf, Solid State Commun. **96**, 1 (1995).
[22] O. Hod, Journal of Chemical Theory and Computation **8**, 1360 (2012).
[23] A. Marini, P. García-González, and A. Rubio, Phys. Rev. Lett. **96,** 136404 (2006).
[24] F. Liu, P. Ming, and J. Li, Phys. Rev. B **76**, 064120 (2007).
[25] Y. Wei, J. Wu, H. Yin, X. Shi, R. Yang, and M. Dresselhaus, Nat. Mater. **11**, 759 (2012).
[26] W. Helfrich, Z. Naturforsch **28c**, 693 (1973).
[27] A. Ramasubramaniam, D. Naveh, and E. Towe, Nano Lett. **11**, 1070 (2011).
[28] M. Han, B. Özyilmaz, Y. Zhang, and P. Kim, Phys. Rev. Lett. **98**, 206805 (2007).
[29] K. Nakada, M. Fujita, G. Dresselhaus, and M. S. Dresselhaus, Phys. Rev. B **54**, 17954 (1996).
[30] L. Brey, and H. Fertig, Phys. Rev. B **73**, 235411 (2006).
[31] Y. W. Son, M. L. Cohen, and S. G. Louie, Phys. Rev. Lett. **97**, 216803 (2006).
[32] Z. H. Chen, Y. M. Lin, M. J. Rooks, and P. Avouris, Physica E **40**, 228 (2007).
[33] V. Pereira, A. Castro Neto, and N. Peres, Phys. Rev. B **80**, 045401 (2009).
[34] F. Guinea, M. I. Katsnelson, and A. K. Geim, Nat. Phys. **6**, 30 (2010).
[35] J. Qi, X. Qian, L. Qi, J. Feng, D. Shi, and J. Li, Nano Lett. **12**, 1224 (2012).
[36] J. Li, G. Gui, and J. Zhong, J. Appl. Phys. **104**, 094311 (2008).
[37] E. Doni, and G. Parravicini, IL Nuovo Cimento B (1965-1970) **64**, 117 (1969).
[38] A. Bhattacharya, S. Bhattacharya, and G. P. Das, Phys. Rev. B **85**, 035415 (2012).
[39] B. Arnaud, S. Lebègue, P. Rabiller, and M. Alouani, Phys. Rev. Lett. **96,** 026402 (2006).
[40] G. Gui, J. Li, and J. Zhong, Phys. Rev. B **78,** 075435(2008).
[41] M. Farjam, and H. Rafii-Tabar, Phys. Rev. B **80,** 167401 (2009).
[42] G. Kresse, and J. Furthmüller, Comput. Mater. Sci. **6**, 15 (1996).
[43] G. Kresse, and J. Furthmüller, Phys. Rev. B **54**, 11169 (1996).
[44] P. E. Blöchl, Phys. Rev. B **50**, 17953 (1994).
[45] G. Kresse, and D. Joubert, Phys. Rev. B **59**, 1758 (1999).
[46] J. P. Perdew, K. Burke, and M. Ernzerhof, Phys. Rev. Lett. **77**, 3865 (1996).
[47] J. P. Perdew, K. Burke, and M. Ernzerhof, Phys. Rev. Lett. **78**, 1396 (1997).
[48] H. J. Monkhorst, and J. D. Pack, Phys. Rev. B **13**, 5188 (1976).

**Tables:**

**Table 1. Mechanical properties of single layer h-BN when mechanical strain is applied in different directions.** Here the strength and the critical strain are the maximum stress and the corresponding strain from the stress-strain curves in Fig. 1b, respectively.

| Loading orientation | Chirality | $\theta$ | Young's modulus (GPa) | Strength (GPa) | Critical strain | Critical bond length (Å) |
|---|---|---|---|---|---|---|
| Zigzag | (1, 0) | 0 | 780±20 | 102 | 0.29 | 1.751 |
| Rotated | (4, 1) | 10.9 | 782±20 | 91 | 0.20 | 1.751 |
| Rotated | (2, 1) | 19.1 | 780±20 | 88 | 0.19 | 1.756 |
| Armchair | (1, 1) | 30 | 773±40 | 88 | 0.18 | 1.760 |
| Bi-axial | / | / | 995±60 | 108 | 0.21 | 1.821 |

**Figures**

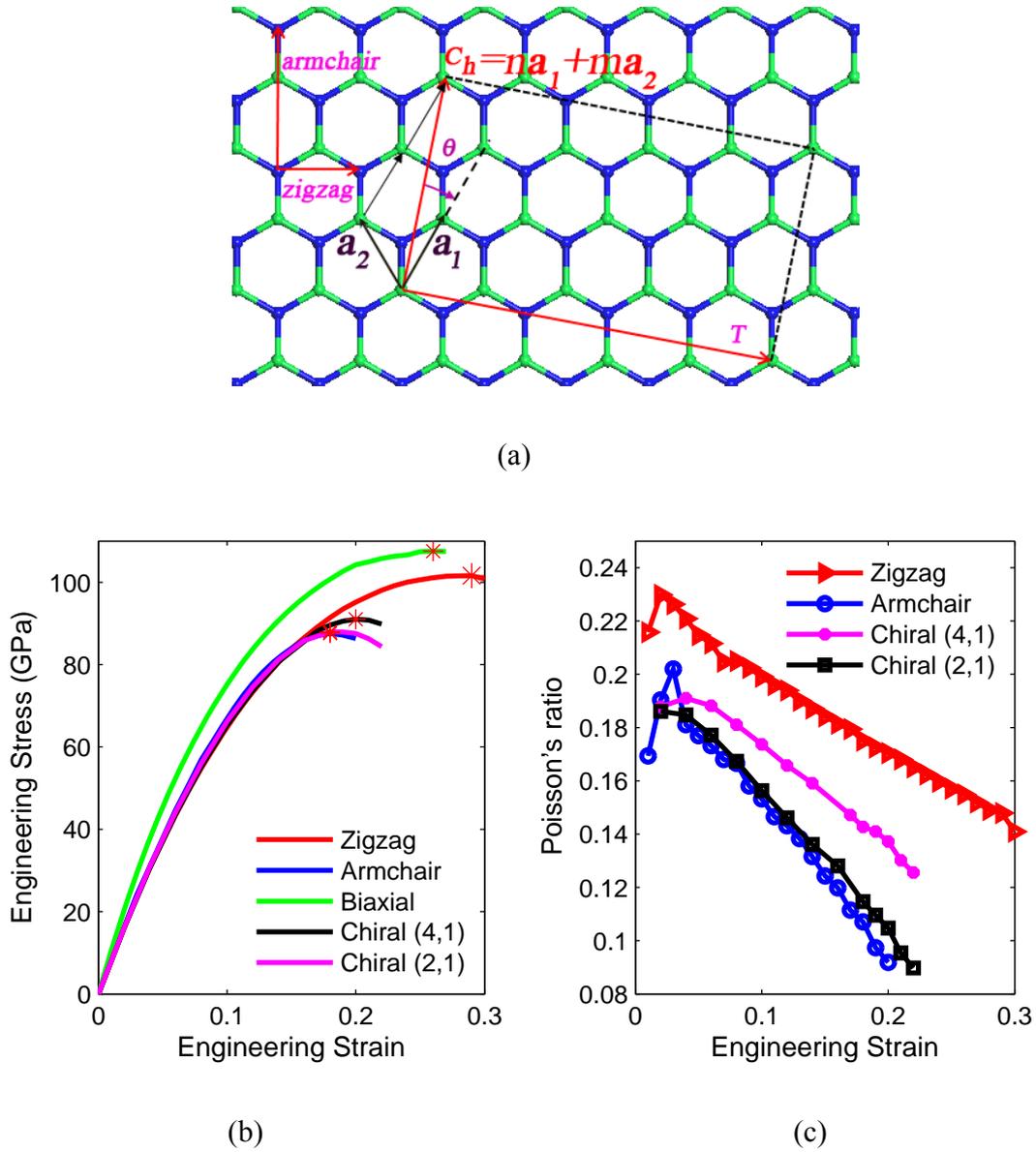

(a)

(b)

(c)

Fig. 1 (Color online): Mechanical behavior of h-BN. (a) Definition of orientations in single layer h-BN. Here $a_1$ and $a_2$ are the unit vectors of h-BN in real space. A typical rectangular sample (with $C_h$=(2,1)) for strain-stress calculations was used. (b) Engineering stress vs. engineering strain curves obtained for samples loaded along different directions. (c) Poisson's ratio as a function of engineering strain for different loading cases as described in the text.

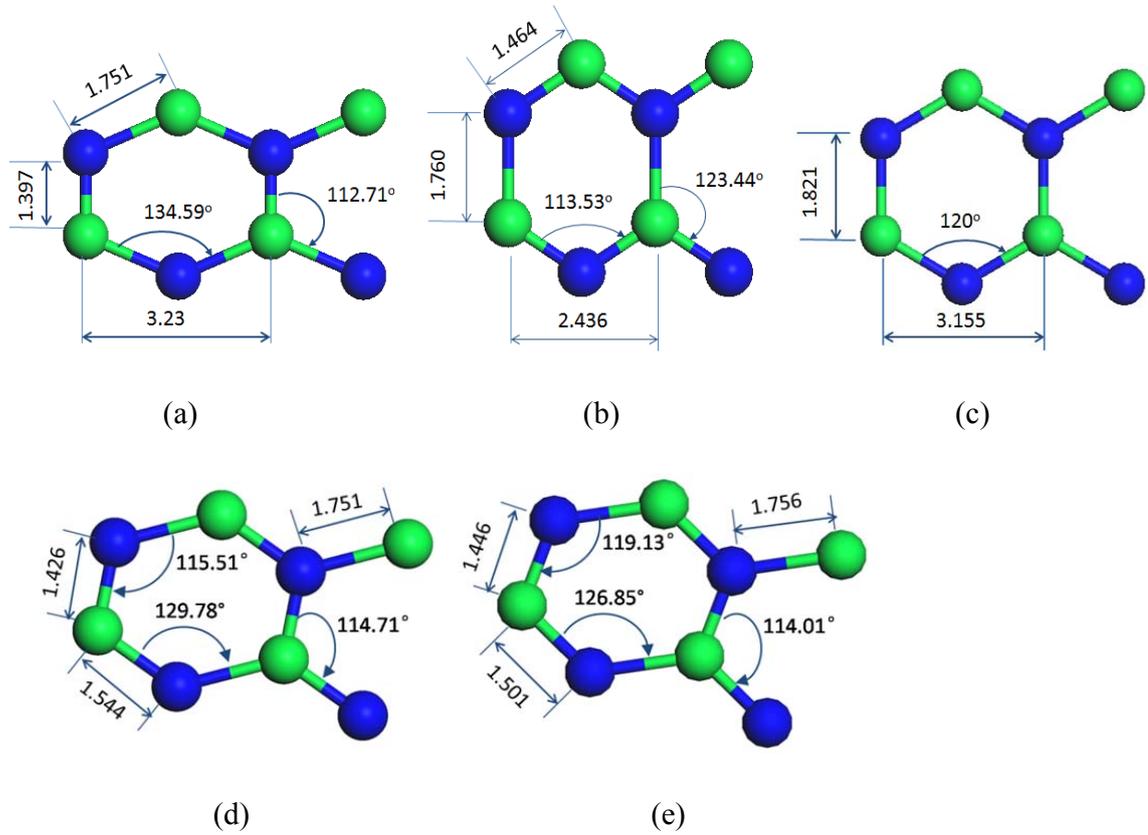

Fig. 2 (Color online): Bond structures and bond-angle information at the failure point where the stress is maximized in the stress-strain curves seen in Fig. 1. (a) Loading along the zigzag direction. (b) Loading along the armchair direction. (c) Biaxial straining. (d) Loading along the direction defined by chirality (4,1), with the angle $\theta$ between the loading direction and zigzag direction $\theta$=10.9º. (e) Loading along the direction defined by chirality (2,1), and $\theta$=19.1º.

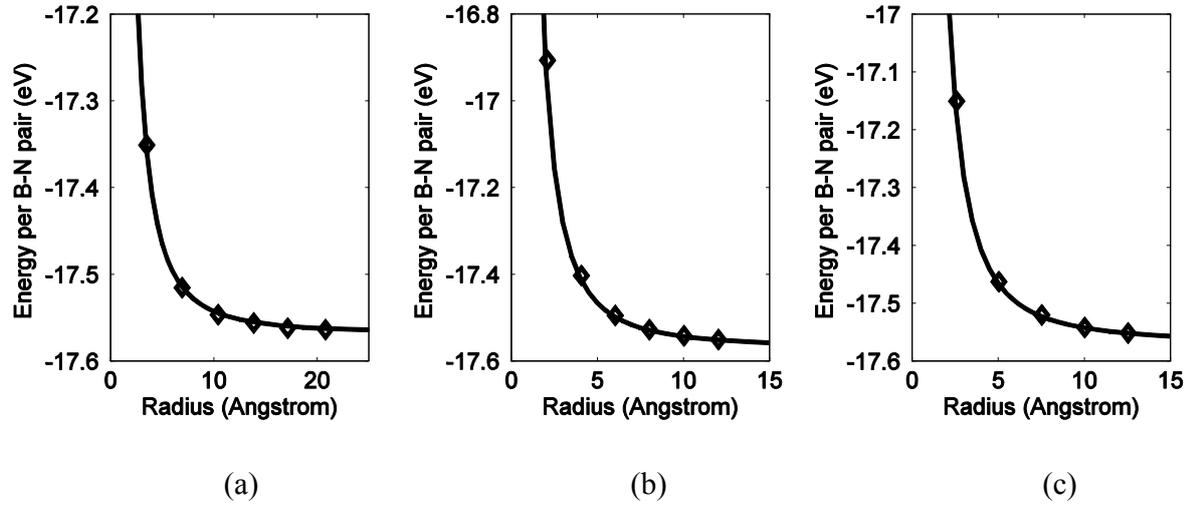

Fig. 3: Energy per B-N pair versus the radius in single-wall h-BN tubes with different rolling up directions: DFT calculations (symbols) and the fitted curve obtained by using eq. (1) with bending rigidity $B_M$=0.95eV (solid lines). (a) For tubes rolled up along the armchair direction. (b) For tubes rolled up along the zigzag direction. (c) For tubes rolled up along the (5,2) chirality.

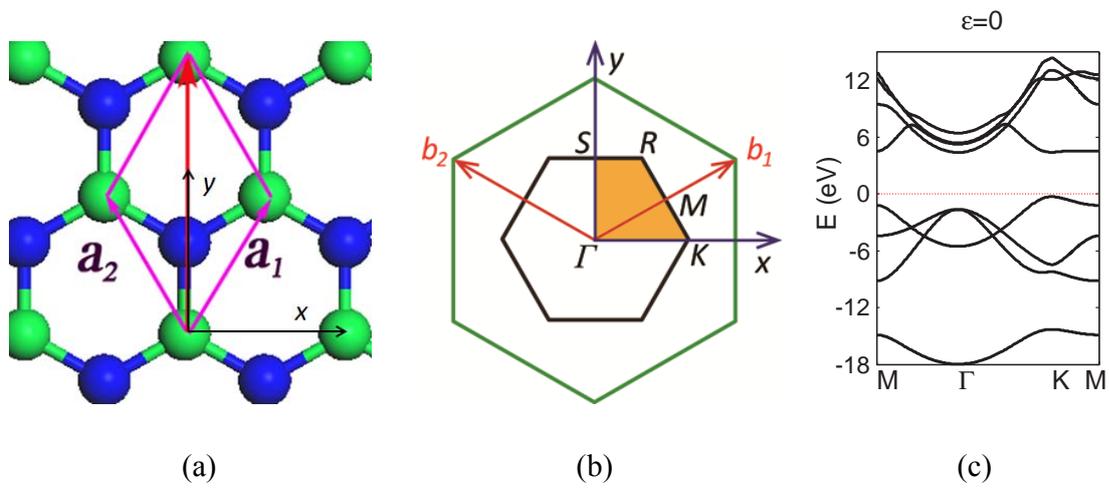

(a) (b) (c)

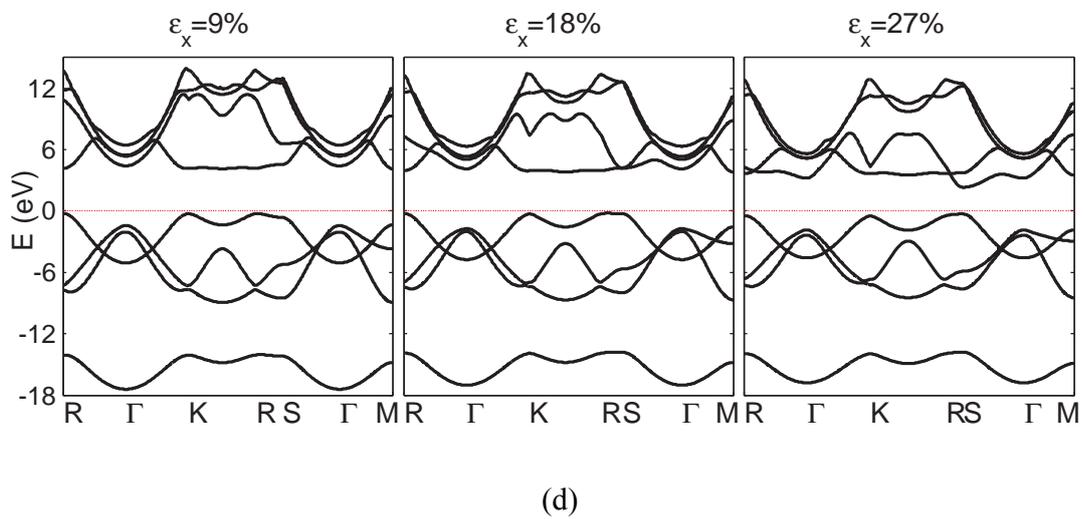

(d)

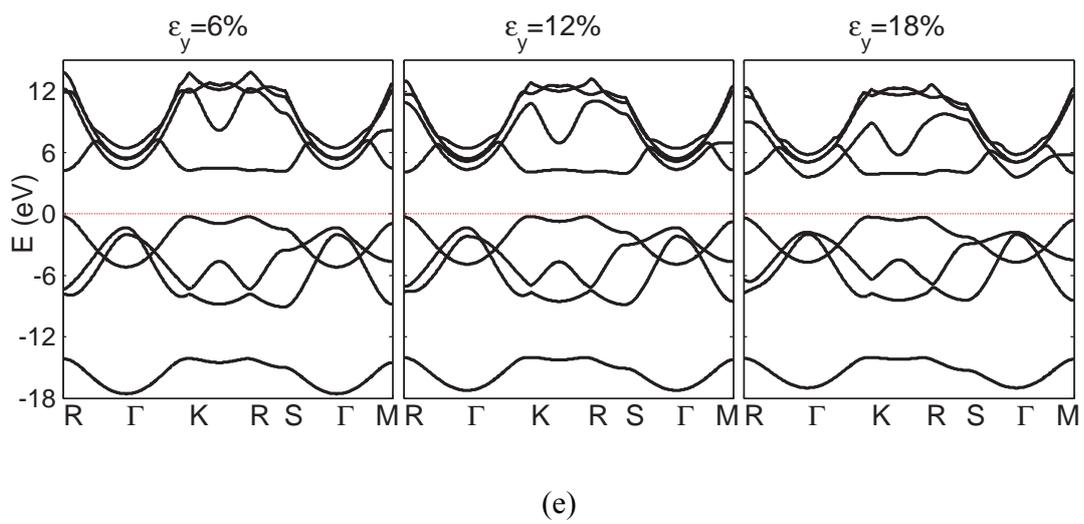

(e)

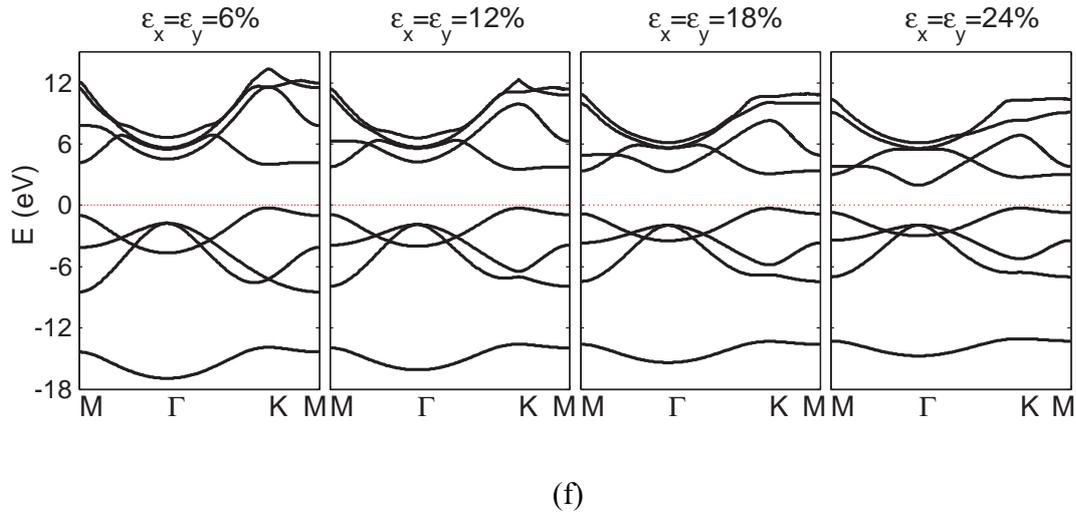

(f)

Fig. 4 (Color online): Band structures of h-BN in response to engineering strains. (a) The two-atom unit cell used for band structure calculations (in the red box). (b) Corresponding reciprocal lattice and the first Brillouin zone. (c) The band structure of monolayer h-BN in the absence of strain. (d) to (f), The evolution of band structures with increasing strains that: (d) strains are applied along the zigzag direction, (e) strains are applied along the armchair direction, (f) strains are bi-axial.

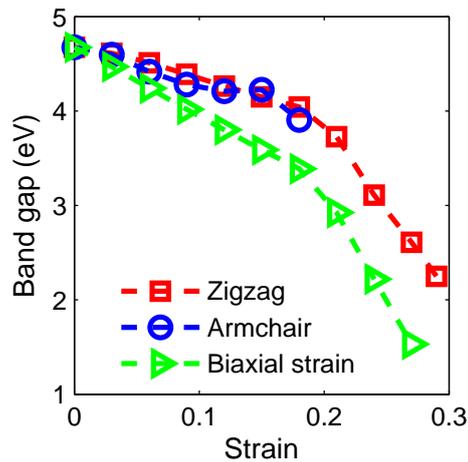

Fig. 5 (Color online): Band gaps as a function of engineering strains for different loading directions.